\documentclass[preprint,prd,showpacs,showkeys,byrevtex,footinbib,eqsecnum,unsortedaddress]{revtex4}
\usepackage{amsfonts}
\usepackage{amsmath}
\usepackage{amssymb}
\usepackage{graphicx}
\input{tcilatex}

\begin{document}

\title{Efficient laser noise reduction by locking to an actively stabilized fiber interferometer with 10 km arm imbalance}
\author{Dawei Li}
\author{Cheng Qian}
\author{Shanglin Li}
\author{Zhengbin Li}
\author{Jianye Zhao}
\email{zhaojianye@pku.edu.cn}
\affiliation{School of Electronics Engineering and Computer Science, Peking University, Beijing, 100871}


\begin{abstract}
We report a laser noise reduction method by locking it to an actively stablized fiber-based Mach–Zehnder interferometer with 10 km optical fiber to achieve large arm imbalance. An acousto-optic modulator is used for interferometer stablization and heterodyne detection. The out-of-loop frequency noise is reduced by more than 90 dB for Fourier frequency at 1 Hz. This structure presents an efficient laser noise reduction method both at high Fourier frequency and low Fourier frequency. The signal of stabilized laser is transferred via a 10 km fiber link with a fractional frequency stability of $1.12 \times 10^{ -16}$ at 1 s. Compared with the fractional frequency stability of that when the interferometer is not stabilized, more than one order of magnitude is improved.
\end{abstract}

\date{\today}
\maketitle

Low phase noise continuous wave lasers are key requirments for applications such as high resolution spectroscopy \cite{de2015use}, optical atomic clocks \cite{jiang2011making}, low noise interferometric sensors \cite{adhikari2014gravitational},  optical frequency transfer via fiber link  \cite{ma1994delivering, newbury2007, terra2010brillouin, fujieda2011, lopez2012, droste2013, calonico2014high, stefani2015tackling} and many others. Erbium-doped fiber distributed-feedback lasers (DFB EDFL)  typically exhibit optical linewidths in the range from 1 kHz to 10 kHz while their performances are still insufficient for many of these applications. 

The phase noise and the linewidth of the continuous wave lasers are usually reduced by the Pound-Drever-Hall (PDH) method \cite{drever1983laser, ludlow2007compact, webster2008thermal, kessler2012sub}. In this method, sub 40 mHz linewidth has been achieved for continuous wave laser \cite{kessler2012sub} and the fractional frequency stability of the stabilized laser has been reduced to $1 \times 10^{-15}$ at 1 s \cite{ludlow2007compact}.  However, this scheme requires fine alignment of free-space optical components, tight polarization adjustment, and spatial mode matching. Moreover, to avoid air-index fluctuations and improve the thermal control and stability, the cavity has to be housed in a high vacuum enclosure with thermal radiation shielding. Therefore, the system is usually bulky, fragile and expensive and it is difficult to adopt this laser stabilization method outside of the laboratory, such as portable interferometric sensors.

An alternative approach to reduce phase noise of the laser is first measuring the phase noise of the laser via a fixed relative time delay line\cite{chen1989use, cranch2002frequency} and then compensating it by a phase locked loop. The fixed relative time delay can be obtained by a two-arm interferometer such as Mach-Zehnder interferometer (MZI). It converts the laser frequency excursion of the optical frequency (${\nu}_{opt}$)  to phase error (${\phi}_{err}(f)$) in radio frequency domain with a transfer function 
\begin{equation}
{\phi}_{err}(f)/{\nu}_{opt}(f)=H_{MZI}(f)=\frac{1-e^{-j2{\pi}f{\tau}}}{jf} (rad/Hz),
\end{equation}
where $\tau$ is the fiber delay time and f is the Fourier frequency. For $f \ll 1/\tau$, $H_{MZI}(f) \approx  2{\pi}{\tau}$. In order to obtain a sufficient frequency discriminator sensitivity, a relatively large arm imbalance is required. Optical fiber is an excellent material to achieve such a large path imbalance. It can lead to a more robust, more compact and cheaper device than the ultrastable cavity. However, the time delay induced by the optical fiber would fluctuate as the optical length of fiber is sensitive to environment-induced acoustical, mechanical and thermal perturbations. These perturbations can significantly degrade the phase noise reduction of the continuous wave laser, expecially at low Fourier frequency. Usually, air-sealed or vacuum chambers with foam are used to passively isolate the interferometer from the influence of the temperature and vibration \cite{sheard2006high, kefelian2009ultralow, jiang2010agile, mcrae2013frequency, mcrae2014digitally, dong2015subhertz}. \\
In this letter, an acousto-optic modulator (AOM) is used to actively compensate the environment induced phase noise in the fiber based MZI which is used to extract the intrinsic frequency excursion of the laser. The scheme is not only portable but also robust to the environment. In our experiment, the arm imbalance of the MZI is 10 km. Its bandwidth is 20 kHz and its quality factor theoretically equals that of a 10 cm Fabry-Perot cavity of which finesse is 330000. When the MZI is actively stabilized, the out-of-loop frequency noise of the stabilized laser is reduced by more than 45 dB at low Fourier frequency compared with that when the MZI has not been stabilized yet.  It means that the frequency noise performance of the laser can be efficiently reduced by this laser stabilization scheme.

Figure 1 shows the  frequency noise reduction scheme of the laser. The phase noise reduction is based on the extracted error signal from a frequency-shifted MZI. The frequency shift MZI converts the optical frequency fluctuation to phase error signal in radio frequency domain. The carrier signal of the phase error is the center frequency of the AOM.  For the MZI in use, its arm imbalance is provided by a 10 km fiber spool. It means that 1 Hz laser frequency excursion will convert into an approximately 314 $\mu$rad phase signal. The input optical wave is splitted between the two arms of MZI by a 50/ 50 fiber coupler (Coupler-1).   The first arm of the MZI consists of an AOM (AOM2) and the fiber spool.  The AOM is used to shift the frequency of the laser signal thereby generating a heterodyne signal between the signals propagating in the two arms. It is also used to compensate the environment induced phase fluctuation in the MZI.  The second arm of the MZI consists of a polarization controller.  By adjusting the polarization controller, the light waves in the two arms of the MZI will share the same state of polarization and lead to a maximum beat-note signal amplitude. From the beat-note, the frequency noise of the laser and the environment-induced phase noise in the interferometer can be extracted.\\

\begin{figure}[htbp]
\centering
\includegraphics[width=0.8\linewidth]{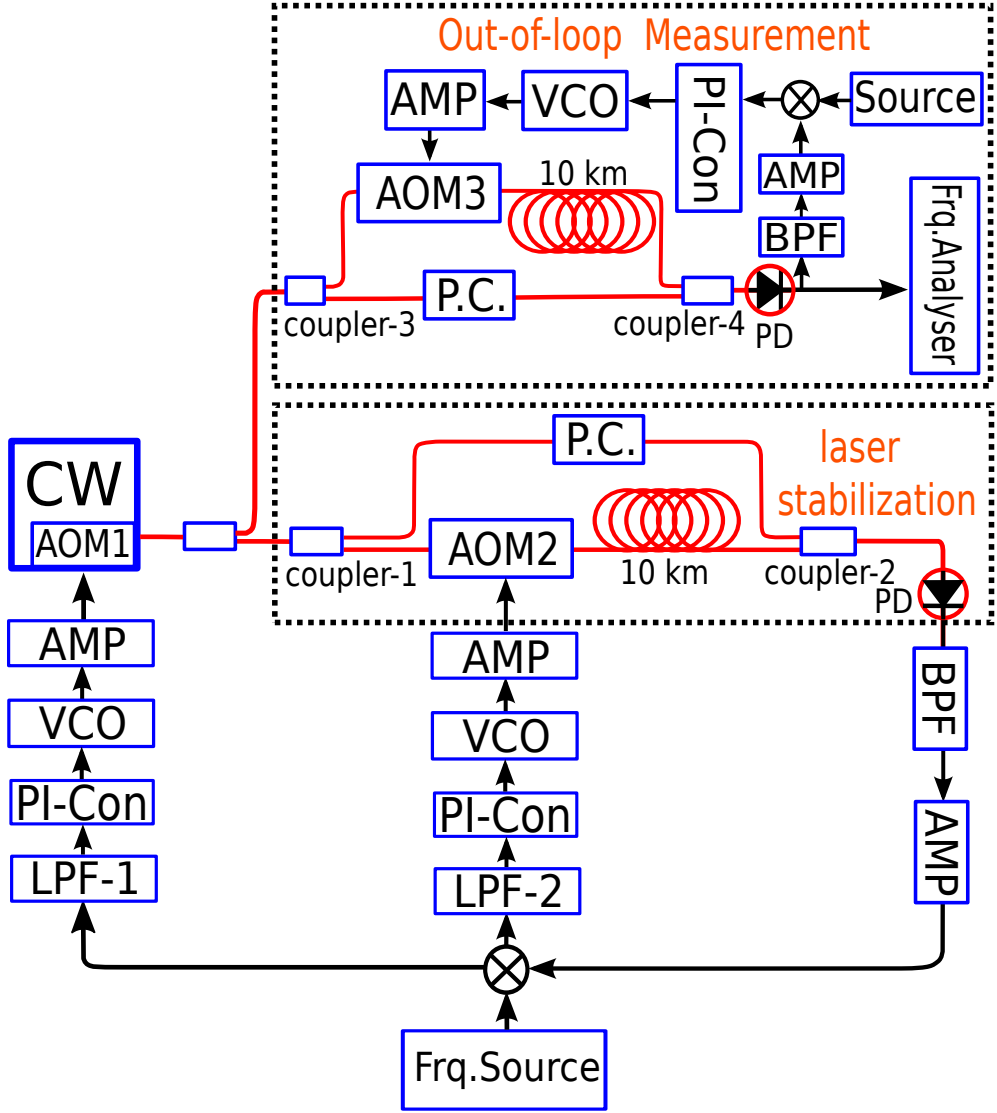}
\caption{ The frequency noise reduction scheme of the CW laser. CW: continuous wave laser, AOM: acousto-optic modulator, AMP: amplifier, VCO: voltage-controlled oscillator, PI-Con: proportional-integral controller, BPF: band pass filter, LPF: low pass filter, Frq.Source: frequency source, Frq. Analyser: frequency analyser, P.C.: polarization controller, PD: photodetector. The couplers are all polarization insensitive. An actively stabilized MZI is used for laser stabilization. In this  MZI, AOM1 is used to stabilize the laser while AOM2 is used to stabilize the interferometer. The signal detected by the photodetector after Coupler-2 is used to stabilize the interferometer and reduce laser noise. Another MZI is used for out-of-loop phase noise measurement of the laser. In this MZI, AOM3 is used to generate a heterodyne between the two interferometer arms. The driving signals of the AOMs are generated by low phase noise voltage-control oscillators and then amplified to the power of 30 dBm. The signal detected by the photodiode after coupler-4 is used to monitor the out-of-loop phase noise of the laser.}
\end{figure}

 The optical power seeded into the interferometer is 3 dBm.  The optical power loss in the first arm is about 12 dB due to the loss of couplers,  AOM2 and the fiber spool. The optical power loss in the second arm is about 7 dB  due to the loss of two couplers and the polarization controller. The beat signal of the two arms is photodetected at the output port of Coulper-2. The detected radio signal contains a carrier signal which is centered at $f_{AOM2}$. This radio signal is phase-modulated by ${\phi}_{laser} + {\phi}_{interferometer} + {\Delta}{\theta}_{rf}$, where ${\Delta}{\theta}_{rf}$ is the local oscillator phase shift, ${\phi}_{laser}$ is the frequency noise of the laser and ${\phi}_{interferometer}$ is the environment-induced phase noise in the MZI. ${\Delta}{\theta}_{rf}$ can be adjusted to zero by changing the phase of the local oscillator signal of the mixer. Then only ${\phi}_{laser}$ and ${\phi}_{interferometer}$ are left. ${\phi}_{interferometer}$ should be reduced by adjusting AOM2 to compensate the environmental influence on the interferometer. Otherwise, it will severely interfere with the frequency noise reduction of the laser. In order to minimize the interference between  ${\phi}_{laser}$ and ${\phi}_{interferometer}$, The laser phase noise reduction and the interferometer stabilization process will be controlled by different proportional-integral controllers (P-I controller). Before the error signal is sent into the P-I controller, it is amplified and filtered in advance. The bandwidth of the low pass filter for the interferometer stabilization loop is below 100 Hz because the environmental influence is predominant in the low Fourier frequencies \cite{kefelian2009ultralow}. The bandwidth of the low pass filter for laser phase noise reduction is 20 kHz due to the bandwidth of the MZI.  However, it is inadequate to achieve a minimum frequency noise for the laser. More importantly, the proportional and integral parameters of the P-I controllers for laser noise reduction and interferometer stabilization should be adjusted together based on the out-of-loop phase noise of the laser. 
 
 \begin{figure}[htbp]
\centering
\includegraphics[width=1\linewidth]{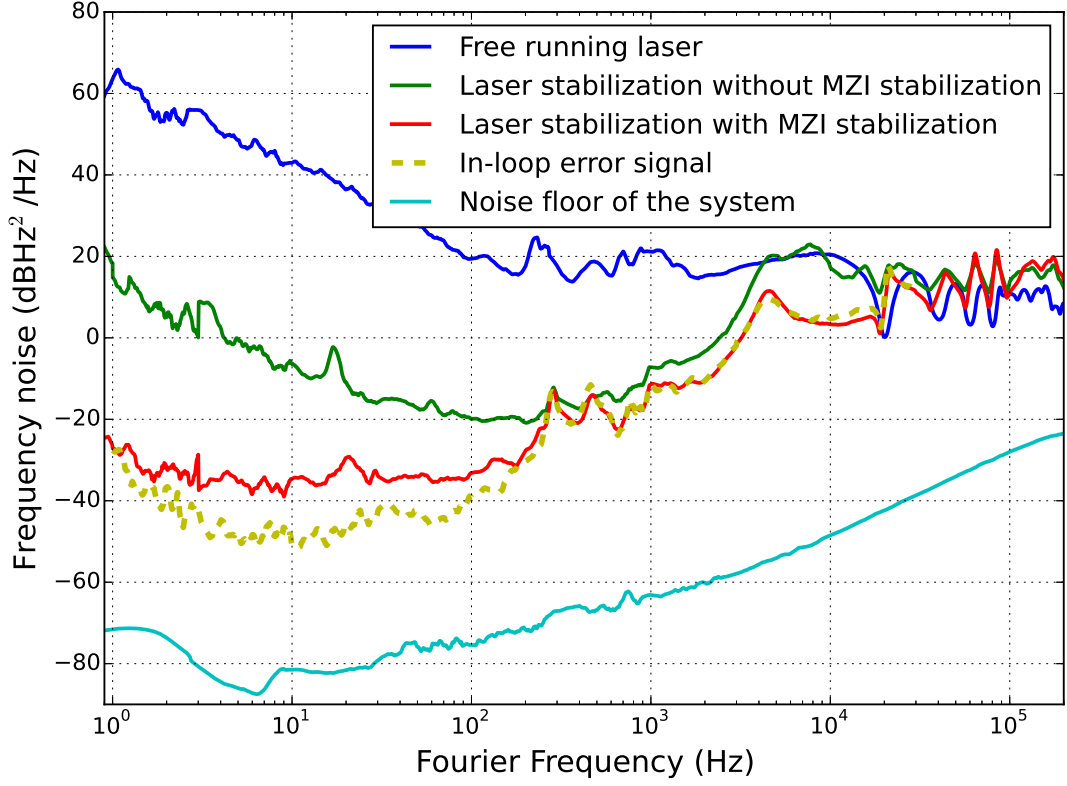}
\caption{The out-of-loop frequency noise of the stabilized laser. The red curve illustrates the frequency noise of the stabilized laser when the MZI is stabilized. The green curve illustrates the frequency noise of the stabilized laser when the MZI is not stabilized. The blue curve illustrates the frequency noise of the free running laser.  The yellow dashed curve illustrates  the in-loop frequency noise of the stabilized laser when the MZI is stabilized. The cyan curve illustrates  the noise floor of the system.}
\end{figure}

The out-of-loop phase noise of the laser is moniterd by another MZI \cite{cranch2002frequency}.  An AOM (AOM3) is used to shift the frequency of the laser signal in an arm of the MZI which is with 10 km fiber spool. In the meantime, it is also used to pre-stabilize the MZI. During the out-of-loop noise measurement, the parameters of this stabilization loop keep unchanged. The optical power photodetected at the output port of Coulper-4 contains a radio frequency which is centered at $f_{AOM3}$.  The signal is phase-modulated by out-of-loop phase noise of the laser. By adjusting the proportional parameters and integral parameters of the laser stabilization loop and its interferometer stabilization loop,  the lowest out-of-loop frequency noise is below -25 dBc/Hz for Fourier frequency at 1 Hz. Fig.2 illustrates the out-of-loop frequency noises of the laser under different states. The frequency noise  is reduced by more than 90 dB  for Fourier frequency around 1 Hz compared with that of the free running laser.  Furthermore,  the phase noise of the laser is reduced by about 45 dB for Fourier frequency at 1 Hz compared with that when the MZI which is used for laser stabilization has not stabilized yet. It illustrates that the environmental influence on the MZI is severe at low Fourier frequency and has been reduced efficiently. However, compared with the in-loop error signal, the frequency noise is still about 10 dB higher  for Fourier frequency from 2 Hz to 200 Hz. This is due to the residual interference between the frequency noise of the laser and environment-induced phase noise of the MZI. Even though we have adjusted the parameters finely, it is still difficult to further minimize the interference.  Due to the spectrum overlapping of intrinsic laser noise and the environment-induced noise, it is not possible to totally eliminate the interference.  \\

In order to test the long-term fractional stability improvement of the laser, we also demonstate an optical frequency transfer experiment. Usually, the  phase noise added by the fiber due to environmental noise dominates the total noise when the laser source is locked to an ultrastable cavity. However, when the laser is not stabilized, the phase noise of the laser might be so large that it should be taken into account \cite{williams2008high}. The contribution from the frequency noise of the laser $S_{Laser}(f)$ to the round-trip phase noise is $4sin^2(2{\pi}f{\tau})S_{Laser}(f)$ which approximately equals $(4{\pi}f{\tau})^2S_{Laser}(f)$.
Where $\tau$ is the transit time for light to propagate over the fiber. The phase noise contribution from the fiber $S_{fiber}(f)$ is ${\alpha}(2{\pi}f{\tau})^2S_{fiber}(f)$, where ${\alpha}$ is for uniform spatial distribution of noise.  The fiber noise is mainly due to the environmental variation such as thermal drift and vibration. Usually, we have
\begin{equation}
S_{Laser}(f) \ll {\alpha}S_{fiber}(f)/4.
\end{equation} 
However, the phase noise compensation scheme for the fiber noise will not work efficiently when the noise level of the laser is high enough because the noise from the laser and the fiber mixes together and can not be distinguished. It means that we can also evaluate the noise level of the laser by testing the frequency transfer performance. Hence we set up an opticial frequency transfer experiment to further examine the noise effect of the laser and  the effect of the laser stabilization scheme. The frequency transfer schematic is illustrated in Fig. 3 \cite{schediwy2013high}.

 The optical-frequency ${\nu}_0$ is sent along a 10 km fiber link to the remote site. Frequency fluctuation ${\Delta}f$ is induced due to vibration and thermal drift. At the remote site, the optical signal passes through an AOM which imparts a frequency shift $f_A+f_C$, where $f_A$ is the center frequency of the AOM and $f_C$ is used to compensate the environmental influence ${\Delta}f$. Therefore, the optical signal after the AOM can be expressed as
\begin{equation}
{\nu}_{1} = {\nu}_0+{\Delta}f+f_A+f_C.
\end{equation}

Part of the signal is detected by a photodetector. Meanwhile, the rest of the signal is reflected by a Faraday mirror (FM), passes through the AOM and propagates over the fiber link for a second time. At the local site, the signal can be expressed as
\begin{equation}
{\nu}_{2} = {\nu}_0+2{\Delta}f+2f_A+2f_C.
\end{equation}
This signal is reflected by another FM and propagates over the link and passes through the AOM for a third time. It can be expressed as
\begin{equation}
{\nu}_{3} = {\nu}_0+3{\Delta}f+3f_A+3f_C.
\end{equation}
Hence the optical power photodetected contains the beat signal of  ${\nu}_{1}$ and ${\nu}_{3}$, and it can be expressed as
\begin{equation}
{f}_{beat} = 2{\Delta}f+2f_A+2f_C.
\end{equation}
\begin{figure}[htbp]
\centering
\includegraphics[width=0.8\linewidth]{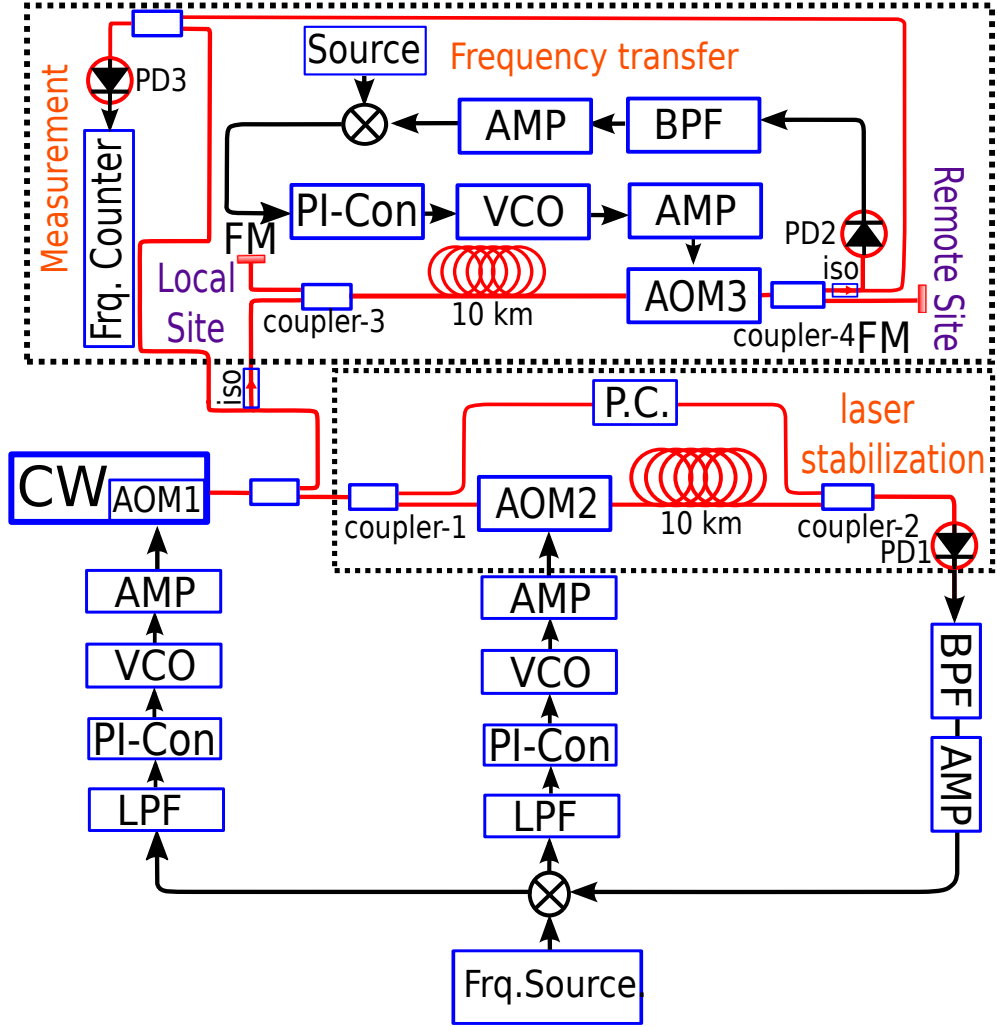}
\caption{The structure of the opticial transfer experiment. CW: continuous wave laser, AOM: acousto-optic modulator, AMP: amplifier, VCO: voltage-controlled oscillator, PI-Con: proportional-integral controller, BPF: band pass filter, LPF: low pass filter, Frq.Source: frequency source, Frq. Analyser: frequency analyser, P.C.: polarization controller, PD: photodetector, FM: Faraday mirror.The couplers are all polarization insensitive. An actively stabilized MZI is used for laser stabilization. In this  MZI, AOM1 is used to stabilize the laser while AOM2 is used to stabilize the interferometer. The signal detected by the photodetector after Coupler-2 is used to stabilize the interferometer and reduce laser noise. A Michelson interferometer (MI) is used for  optical frequency transfer in which local and remote sites are situated in the same laboratory. In this MI, AOM3 is used to stabilize the 10 km fiber spool. The signal detected by the photodiode after coupler-4 is used to measure the frequency fluctuation thereby analyzing the fractional stability of the transfer.}
\end{figure}

The signal ${f}_{beat}$ is bandpass-filtered and then downconverted by a rf mixer driven by a low-noise reference oscillator (R$\&$S SMB100A) at $2f_A$. The output signal $2{\Delta}f+2f_C$ is kept zero to stabilize the optical fiber link. $f_{beat}$ is also measured by a frequency counter (Agilent 53230A) to evaluate the transfer performance. $f_{beat}$ is 100 MHz in this frequency transfer system. Mode "CONT" of the frequency counter is used to record the frequency data of the 100 MHz beat frequency signal with a measurement bandwidth of 1 Hz. The fractional frequency stability of the stabilized optical fiber link is shown in Fig. 4. After the phase noise of the laser is reduced with the stabilization of the MZI, the allan deviation is $1.12 \times 10^{-16}$ at 1s and reaches $3 \times 10^{-21}$ at 30000s. This performance meets the requirement of transferring the most precise optical clock nowadays. When the phase noise of the laser is reduced without compensating the environment-induced phase noise in the MZI, the allan deviation is $5 \times 10^{-15}$ at 1s. It indicates that the short-term fractional stability is more than one order magnitude better due to the stabilization of MZI.  When the laser is in its free running state, the allan deviation is $6.5 \times 10^{-15}$ at 1s. It means that the environment-induced phase noise in the MZI predominates at low Fourier frequency. 

\begin{figure}[htbp]
\centering
\includegraphics[width=1\linewidth]{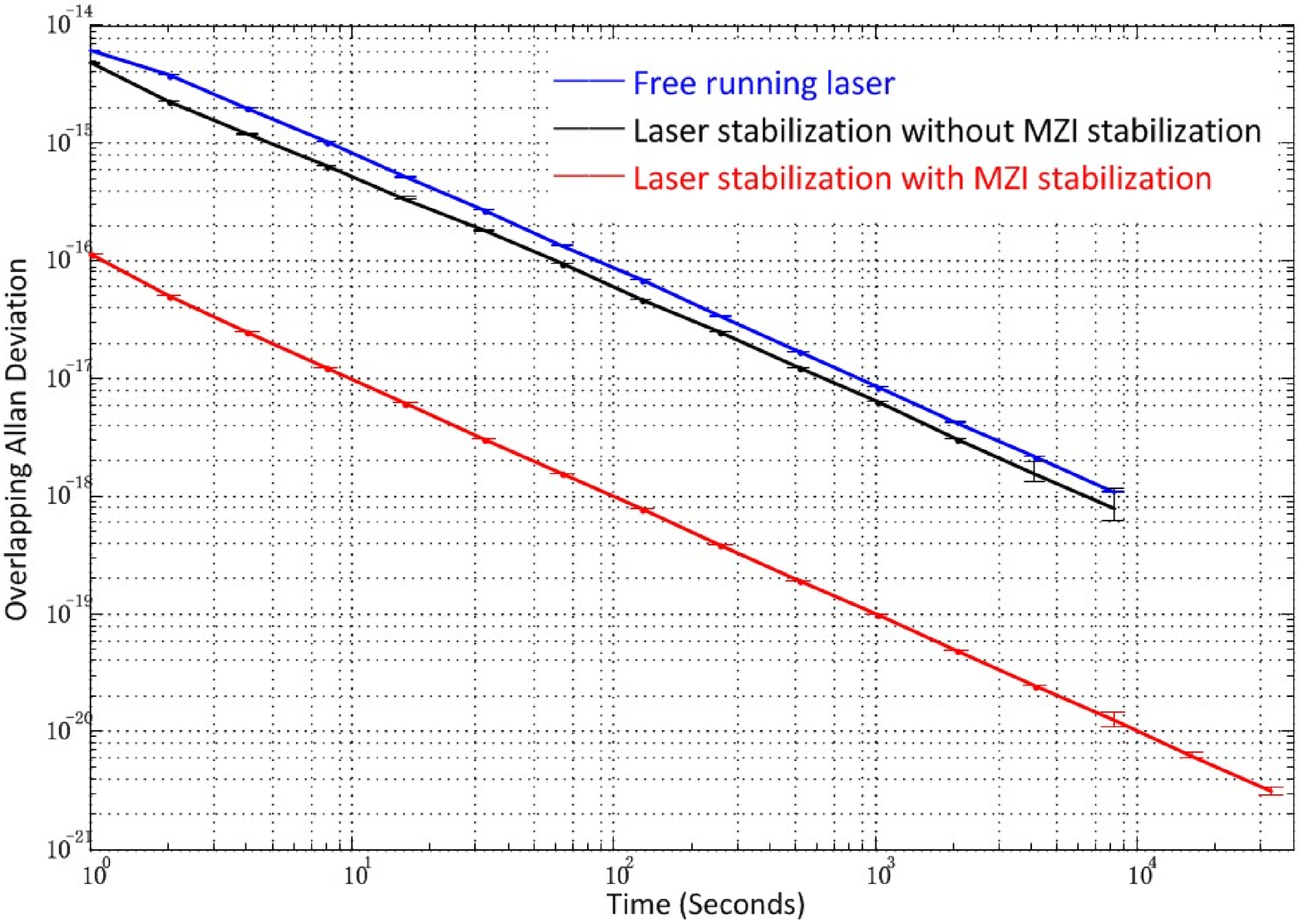}
\caption{The allan deviation of the optical transfer experiment. The fractional stability when the phase noise of the laser is reduced with the stabilization of the MZI (red curve) versus the fractional stability when the phase noise of the laser is reduced without the stabilization of the MZI (black curve) and the fractional stability when the  laser is at its free running state (blue curve)}
\end{figure}
In conclusion, we have demonstrated a  noise reduction method of CW laser by locking it to an actively stabilized  MZI with 10 km arm imbalance. By actively stabilizing the MZI, the frequency noise of the laser is reduced significantly, especially at low Fourier frequency.  The frequency noise is about 45 dB lower compared the result when the laser is stabilized without stabilizating the MZI at low Fourier frequency. This improvement is  due to the efficient active reduction against the  environment-induced phase noise to the MZI.  The influences used to be isolated by air-sealed or vacuum chambers with foam or other passive protection method. However, these methods make the laser stabilization scheme bulky. Compared with these methods, our scheme is both portable and less immune to the environmental influence. It provides a more compact, robust, and flexible alternative to the ultra-stable cavity locking method with an all-fiber system.  \\

National Science Foundation of China (NSFC) (61371074).\\

The authors thank Yunfeng Zhang and Yaolin Zhang for technical discussion.

\end{document}